\documentclass[%
 reprint,
nofootinbib,
 amsmath,amssymb,
 aps,
 prd,
floatfix,
]{revtex4-2}

\usepackage{csquotes}
\usepackage{nccmath}
\usepackage{physics}
\usepackage{slashed}
\usepackage{nicefrac}
\usepackage{graphics}
\usepackage{graphicx}
\usepackage[table]{xcolor}
\usepackage{rotating}
\usepackage{dcolumn}    % Align table columns on decimal point
\usepackage{bm}
\usepackage[T1]{fontenc} 
\usepackage{bbold}
\usepackage{dsfont}
\usepackage{multirow}
\usepackage{tikz}
\usepackage{url}
\usepackage{hyperref}  % LAST package
\hypersetup{
    colorlinks=true,
    citecolor=[rgb]{0,0.334,0.666}, %{0,0.482,0.655}
    linkcolor=[rgb]{0,0.334,0.666},
    urlcolor=[rgb]{0,0.334,0.666},
}

\DeclareMathOperator{\0}{\mathbb{0}}
\DeclareMathOperator{\1}{\mathbb{1}}

\newcommand{\Var}{\operatorname{Var}} 
\makeatletter
\DeclareRobustCommand{\cev}[1]{%
  {\mathpalette\do@cev{#1}}%
}
\newcommand{\do@cev}[2]{%
  \vbox{\offinterlineskip
    \sbox\z@{$\m@th#1 x$}%
    \ialign{##\cr
      \hidewidth\reflectbox{$\m@th#1\vec{}\mkern4mu$}\hidewidth\cr
      \noalign{\kern-\ht\z@}
      $\m@th#1#2$\cr
    }%
  }%
}
\makeatother

\begin{document}

\title{Renormalization footprints in the phase diagram of the Grosse-Wulkenhaar model} 

\author{Dragan Prekrat}
\email{dprekrat@ipb.ac.rs}
\affiliation{University of Belgrade, Faculty of Physics, P.O. Box 44, SR-11001 Belgrade, Serbia}

\date{\today}

\begin{abstract}
We construct and analyze the phase diagram of a self-interacting matrix field in two dimensions coupled to the curvature of the non-commutative truncated Heisenberg space. In the infinite size limit, the model reduces to the renormalizable Grosse-Wulkenhaar's. The curvature term proves crucial for the diagram's structure: when turned off, the triple point collapses into the origin as matrices grow larger; when turned on, the triple point recedes from the origin proportionally to the coupling strength and the matrix size. The coupling attenuation that turns the Grosse-Wulkenhaar model into a renormalizable version of the $\phi^4_\star$-model cannot stop the triple point recession. As a result, the stripe phase escapes to infinity, removing the problems with UV/IR mixing. 
\end{abstract}

                              
\maketitle

%\tableofcontents

%%%%%%%%%%%%%%%%%%%%%%%%%%%%%%%%%%%%%%%%%%%%%%%%%%%%%%%%%%%%%%%%%%%%%%%%%%%%%%%%%%%%%
%%%%%%%%%%%%%%%%%%%%%%%%%%%%%%%%%%%%%%%%%%%%%%%%%%%%%%%%%%%%%%%%%%%%%%%%%%%%%%%%%%%%%
%%%%%%%%%%%%%%%%%%%%%%%%%%%%%%%%%%%%%%%%%%%%%%%%%%%%%%%%%%%%%%%%%%%%%%%%%%%%%%%%%%%%%

\section{Introduction}
\label{sec:intro}

Non-commutativity (NC) of space-time coordinates was initially proposed in the 1940s in the hope of resolving the confusion about the infinities in the nascent quantum field theory \cite{PhysRev.71.38}. The first promising results in this regard were, however, achieved by the technique of renormalization. Since then, NC occasionally reemerged, both in the fundamental and the effective form, from condensed matter physics to quantum gravity \cite{NC2,NC3}. Finally, when NC was discovered in the low energy sector of the string theory at the turn of the millennium \cite{Seiberg:1999vs}, various new NC models followed.  

Contrary to the expected better-than-commutative behaviour, NC models experience more difficulties with renormalizability. Generically, their non-planar Feynman diagrams entangle small and large length scales, which prevents a successful absorption of divergences into the action terms \cite{Minwalla:1999px,Chu:2001xi,Dolan:2001gn,martin2020uvir}. It was shown that this UV/IR mixing could be resolved by the proper balancing of the scales provided by the Langman-Szabo duality \cite{Langmann_2002}. 

Grosse-Wulkenhaar (GW) model \cite{Grosse:2003nw,Disertori:2006nq,Wang:2018sed}
\begin{equation}
S_\text{GW} = \int
\frac{1}{2}(\partial\phi)^2
+ \frac{\Omega^2}{2}((\theta^{-1}x)\phi)^2
+ \frac{m^2}{2}\phi^2
+ \frac{\lambda}{4!}\phi^4
\label{GWmodel}
\end{equation}
managed to evade the UV/IR mixing problem. It features a self-interacting real scalar field on the Moyal space equipped with a $\star$-product
\begin{equation}
    f \star g= f\,e^{\nicefrac{i}{2}\,\cev{\partial}\theta\vec{\partial}}\,g
\quad\Rightarrow\quad
    \comm{x^\mu}{x^\nu}_\star=i\theta^{\mu\nu}.
\end{equation}
Its potential is enhanced by the external harmonic oscillator term of a possible gravitational origin. Namely, the model can be reinterpreted \cite{Buric:2009ss} as that of a scalar field in a curved NC space of the truncated Heisenberg algebra $\mathfrak{h}^\text{tr}$:
\begin{equation}
S_\mathfrak{h} = \int \!\sqrt{g}
\left(
\frac{1}{2}(\partial\phi)^2
- \frac{\xi}{2}R_\mathfrak{h}\phi^2
+ \frac{m^2_\mathfrak{h}}{2}\phi^2
+ \frac{\lambda_\mathfrak{h}}{4!}\phi^4
\right).
\label{Sr}
\end{equation}
The oscillator $\Omega$-term, which holds the key to renormalizability, is now seen as a coupling to the coordinate-dependant curvature $R_\mathfrak{h}$. More details on the $\mathfrak{h}^\text{tr}$ and $S_\mathfrak{h}$ will be provided later in the text. Another possible source of the oscillator term was presented in \cite{Franchino-Vinas:2019nqy}, where it elegantly appears in the expansion of the kinetic term of the free scalar field situated in the Snyder-de Sitter space. This model also predicts the running of the curvature coupling, which is an essential ingredient of the GW-mediated $\phi^4_\star$-renormalizability. It would be interesting to see if similar conclusions could be reached in the fuzzy de Sitter space \cite{Buric:2017yes,Buric:2019yau}.

UV/IR mixing still poses a problem for gauge fields on NC spaces \cite{Blaschke:2014vda}. Hoping to build on the GW model's success, \cite{Buric:2010xs,Buric:2012bb} tried to adapt it to a gauge field on $\mathfrak{h}^\text{tr}$. Still, after extensive treatment, we found non-renormalizability lurking in the form of divergent non-local derivative counterterms \cite{Buric:2016lly}. It turned out that, apart from the trivial vacuum, this model contains another, which breaks the translational invariance. This echoes the translational symmetry-breaking stripe phase that seems to be at the root of UV/IR mixing. Its \enquote{stripes} refer to patterns of spatially non-uniform magnetization, which appear when the field oscillates around different values at different points in space \cite{Gubser:2000cd,Castorina:2007za,Mejia-Diaz:2014lza}. They also seem to shatter the symmetry between large and small scales that keeps the UV/IR mixing in check: locally, vacuum appears ordered, but globally, watched through the lenses of spatial-averaging, it looks smudged into a disordered zero. It would be interesting to find out what happens with the stripe phase in the GW model. We would like to see how its renormalizability plays out from the phase transition point of view.  

Phase diagrams on NC spaces have been extensively studied in various matrix models, since they regularize corresponding continuum theories in a numerical simulation-friendly fashion \cite{Gross:1980he,Martin:2004un,Panero:2006bx,OConnor:2007ibg,GarciaFlores:2009hf,Lizzi:2012xy,Polychronakos:2013nca,Tekel:2014bta,Ydri:2014rea,Rea:2015wta,Tekel:2015uza,Tekel:2015zga,Ydri:2015vba,Ydri:2016dmy,Sabella-Garnier:2017svs,Tekel:2017nzf,Hatakeyama:2018qjr,Kovacik:2018thy,Prekrat:2020ptq,Subjakova:2020prh,Subjakova:2020shi}. They generically feature three phases that meet at a triple point. Two of these are readily present in commutative theories: in the disordered phase, field eigenvalues clump around zero, and in the ordered phase around one of the mirror image-minima of the potential. The third one is a matrix counterpart of the NC stripe phase: eigenvalues there gather both around positive and negative minimum at the same time. 

In \cite{Prekrat:2020ptq}, we started a numerical comparison of the two-dimensional GW-model matrix regularization with ($R$-on) and without ($R$-off) the curvature term, focusing mainly on the latter. Here, we wish to present more details on the former and to see how it bares under the oscillator term switching off procedure that ensures the $\phi^4_\star$-model's renormalizability \cite{Grosse:2003nw}. Moreover, since the triple point controls the extension of the problematic stripe phase, we will try to pinpoint its location.

The paper is organized as follows. We first reintroduce the model and present its detailed $N=24$ phase diagrams. Then we track the $R$-off triple point as we increase the matrix size. Finally, we present the effects of the curvature coupling variation on the phase diagram, look at the coupled model as we turn the coupling off, and compare its limit with the uncoupled one.

%%%%%%%%%%%%%%%%%%%%%%%%%%%%%%%%%%%%%%%%%%%%%%%%%%%%%%%%%%%%%%%%%%%%%%%%%%%%%%%%%%%%%
%%%%%%%%%%%%%%%%%%%%%%%%%%%%%%%%%%%%%%%%%%%%%%%%%%%%%%%%%%%%%%%%%%%%%%%%%%%%%%%%%%%%%
%%%%%%%%%%%%%%%%%%%%%%%%%%%%%%%%%%%%%%%%%%%%%%%%%%%%%%%%%%%%%%%%%%%%%%%%%%%%%%%%%%%%%

\section{Matrix model}

We here continue inspection of the matrix regularization $S_N$ of \eqref{Sr} started in \cite{Prekrat:2020ptq}. Let us reintroduce the model and walk through its main features. 

The coordinates $x$, $y$ and $z$ of the underlying $\mathfrak{h}^\text{tr}$ algebra satisfy
\begin{subequations}
%\vspace{-5pt}
\begin{equation}
\comm{\mu x}{\mu y} = i\epsilon(1 - \mu z),
\end{equation}
\vspace{-23pt}
\begin{equation}
\comm{x}{z} = +i\epsilon\acomm{y}{z},
\qquad
\comm{y}{z} = -i\epsilon\acomm{x}{z}, 
\end{equation}
\label{comm-rel}
\end{subequations}
\!\!where $\mu$ represents the mass scale and $\epsilon$ the strength of NC. If we set $\epsilon=1$, $\mu x$ and $\mu y$ can be represented by finitely-truncated matrices of the Heisenberg algebra in the energy basis of the harmonic oscillator 
\begin{equation}
\!\!\!\!\!\!\!\!
X =
\frac{1}{\sqrt{2}}\spmqty{
& \sqrt{1} \\
\sqrt{1} &  & \sqrt{2}\\
 & \sqrt{2} &  & \\
 &  & \begin{rotate}{5}$\smash{\scriptsize{\ddots}}$\end{rotate}\, 
}_{_{\!\!\!\!N\times N}}\!\!\!\!\!\!\!\!,
\quad\;
Y =
\frac{i}{\sqrt{2}}\spmqty{
& \!\!\text{-}\sqrt{1} \\
\sqrt{1} &  & \!\!\text{-}\sqrt{2}\\
 & \sqrt{2} &  & \\
 &  & \begin{rotate}{5}$\smash{\scriptsize{\ddots}}$\end{rotate}\,
}_{_{\!\!\!\!N\times N}}\!\!\!\!\!\!\!\!,
\label{XY}
\end{equation}
Identification \cite{Buric:2009ss} of $S_\text{GW}$ and $S_\mathfrak{h}$ requires restriction to subspace $z=0$, which happens in a weak $N\to\infty$ limit of the matrix representation. This turns \eqref{comm-rel} into the Moyal space commutation relations. 

The space $\mathfrak{h}^\text{tr}$ has a curvature
\begin{equation}
    R_\mathfrak{h} = \frac{15\mu^2}{2} - 4\epsilon\mu^3 z - 8\epsilon^2\mu^4(x^2+y^2),
\end{equation}
which is represented (on $z=0$) by a negative diagonal matrix $R$ with elements
\begin{equation}
    R_{ii}=\frac{31}{2}-
\begin{cases}
16i, & \!\!i<N, \\
8N, & \!\!i=N.
\end{cases}\;\,
\end{equation}
In its eigenvalues, we recognize energy levels of the harmonic oscillator. The quadratic dependence of $R_\mathfrak{h}$ on coordinates is precisely what we have in $\Omega$-term in \eqref{GWmodel}.

Derivatives in model \eqref{Sr}, analysed in the frame formalism, are realised as commutators $\partial_\mu = \comm{p_\mu}{\cdot\;}$ with momenta $p_\mu$ 
\begin{equation}
    \frac{\epsilon p_1}{i\mu} = +\mu y,
    \qquad
    \frac{\epsilon p_2}{i\mu} = -\mu x,
    \qquad
    \frac{\epsilon p_3}{i\mu} = \mu z- \frac{1}{2},
\end{equation}
hence their matrix counterparts are
\begin{equation}
P_{1}=-Y,
\qquad
P_{2}=X,
\qquad
P_{3} = \frac{\1}{2}.
\end{equation}

The numbered identifications associate \eqref{Sr} with the matrix model
\begin{equation}
S_N = \Tr\left(
\Phi\mathcal{K}\Phi
- c_r R\Phi^2
- c_2\Phi^2 + c_4\Phi^4
\right),
\label{S_N}
\end{equation}
in which the field $\Phi$ is a $N\times N$ hermitian matrix and $\mathcal{K}$ the kinetic operator 
\begin{equation}
\mathcal{K}\Phi = \comm{P_\alpha}{\comm{P_\alpha}{\Phi}}.
\end{equation}
All originally dimensionful quantities are now expressed in units of $\mu$. We chose the minus sign of the mass term to enable positive $c_2$ to parameterize the relevant portion of the phase diagram, while positive $c_4$ ensures that $S_N$ is bounded from below. They will be accompanied by the rescaled model parameters
\begin{equation}
    \widetilde{c}_2=\frac{c_2}{N},
    \qquad
    \widetilde{c}_4=\frac{c_4}{N}.
\end{equation}

We performed parallel hybrid Monte Carlo simulations to measure various thermodynamic observables, most important being
\begin{itemize}
    \item heat capacity % per degree of freedom 
    $C=\Var S\big/N^2$,
    \item magnetic susceptibility % per eigenvalue 
    $\chi=\Var\abs{\Tr\,\Phi}\,\big/N$,
    \item distributions of eigenvalues and traces of $\Phi$,
\end{itemize}
where expectation value $\expval{\mathcal{O}}$ and variance $\Var\mathcal{O}$ of the observable $\mathcal{O}$ are given by
\begin{equation}
    \expval{\mathcal{O}} = \frac{\int \dd{\Phi} \mathcal{O} \, e^{-S}}{\int \dd{\Phi} e^{-S}},
    \quad
    \;
    \Var\mathcal{O}=\expval{\mathcal{O}^2}-\expval{\mathcal{O}}^2.
    \label{observable}
\end{equation}
Phase transitions in finite systems form smeared finite peaks and edges in profiles of free energy derivatives. Different quantities yield slightly different estimates of transition points, but they ultimately converge for large enough matrices. To locate them, we scanned through parameter space by varying $c_2$ at fixed $c_4$, which played a role of the temperature, and searched for peaks in $C$ and $\chi$.  

\begin{figure}[t]
\centering  
\includegraphics[scale=0.55]{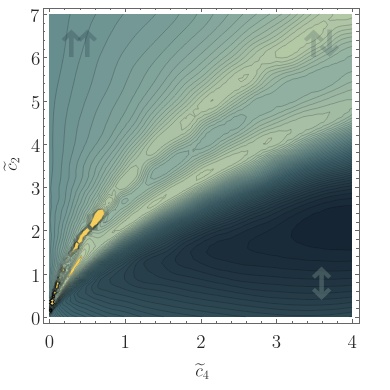}
\includegraphics[scale=0.55]{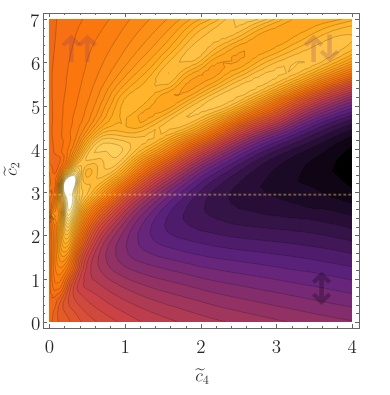}
\caption{Contour plots of $N=24$ phase diagram for $c_r=0$ (top) and for $c_r=0.2$ (bottom). Darker colors depict lower and lighter colors higher values of specific heat, bright stripes being the transition lines. The dotted line on the bottom plot indicates a diagram shift relative to the $R$-off case. Phases are denoted by semi-transparent arrows in the corners of the plot: $\updownarrow$-phase occupies the bottom right, $\uparrow\uparrow$-phase the upper left, while the $\uparrow\downarrow$-phase is sandwiched in between, extending towards the upper right corner. Diagrams are constructed based on more than $5500$ points.
}
\label{diagrams}
\end{figure}

Classical equation of motion for $S_N$
\begin{equation}
2\mathcal{K}\Phi
-c_r\acomm{R}{\Phi}
+ \Phi\left(-2c_2+4c_4\Phi^2\right) = 0,
\label{EOM1}
\end{equation}
gives us an idea what kind of phases to expect. Its kinetic, curvature and pure potential parts are respectively solved by
\begin{equation}
\Phi \propto \1,
\qquad
\Phi = \0,
\qquad
\Phi^2 = 
\begin{cases}
\hspace{5pt} \0, & c_2 \leq 0, \\
\displaystyle \frac{c_2\1}{2c_4}, & c_2 > 0,
\end{cases}
\label{EOM2}
\end{equation}
corresponding to three phases depicted\footnote{Throughout this text, we will use \emph{Wolfram Mathematica} bluish \emph{StarryNightColors} scheme for $R$-off plots and reddish \emph{SunsetColors} scheme for $R$-on plots.} in FIG. \ref{diagrams}:
\begin{itemize}
\item disordered $\updownarrow$-phase: $\expval{\Phi}_{\updownarrow} = \0$,
\item uniformly ordered $\uparrow\uparrow$-phase: $\expval{\Phi}_{\uparrow\uparrow} \propto \1$,
\item non-uniformly ordered $\uparrow\downarrow$-phase: $\expval{\Phi}_{\uparrow\downarrow} \propto \, U\1_\pm U^\dag$, where $UU^\dag=U^\dag U=\1$, $\1_\pm^2=\1$ and $\abs{\Tr\1_\pm}<N$.
\end{itemize}
The $\uparrow\downarrow$-phase is a matrix equivalent of the stripe phase.
Large mass parameter lives in $\uparrow\uparrow$-phase, and large quartic coupling in $\updownarrow$-phase, with $\uparrow\downarrow$-phase nested in between. The phases meet at a triple point.

When the kinetic term is negligible (e.g. field near $\propto \1$) and $c_2 \geq \max_i\{c_r \abs{R_{ii}}\}$, a diagonal solution exists that combines the effects of the curvature and the potential and which deforms the vacuum of the ordered phases:
\begin{equation}
\Phi^2 = \frac{c_2\1+c_r R}{2c_4}.
\label{phiR}
\end{equation}

Both transitions out of the stripe phase in the $R$-off case (top plot in FIG. \ref{diagrams}) follow the square root behaviour for larger quartic coupling. For $N=24$ and $\widetilde{c}_4>1$, they are well approximated by:
%\vspace{-12pt}
\begin{subequations}
\begin{eqnarray}
    \updownarrow \to \uparrow\downarrow: \quad \widetilde{c}_2&=&2.67(5)\sqrt{\widetilde{c}_4}-0.55(7),\qquad
\\
    \uparrow\downarrow \to \uparrow\uparrow: \quad \widetilde{c}_2&=&3.99(4)\sqrt{\widetilde{c}_4}-0.90(5).\qquad
\end{eqnarray}
\end{subequations}
For comparison, a pure potential model would show only a $\updownarrow \to \uparrow\downarrow$ transition line $\widetilde{c}_2=2\sqrt{\widetilde{c}_4}$ in the infinite $N$ limit.

In \cite{Prekrat:2020ptq} we presented a simple argument that, due to diagonality, curvature acts as a quasi-mass term and should shift transition lines by $\delta c_2$ relative to the $R$-off case. The shift $\delta c_2$  is proportional to the $c_r$, and their ratio is bounded by:
\begin{equation}
\frac{1}{2N}
 \le \frac{\delta\hspace{0.5pt}\widetilde{c}_2}{c_r} \le
16-\frac{63}{2N}.
\label{shift}
\end{equation}
We previously demonstrated this by numerical simulation at a token value of quartic coupling and with the absent kinetic term; here, we expose this effect in full in FIG. \ref{diagrams}. Similar shifting is in the meantime also reported on the fuzzy sphere after adding a modification to the kinetic term \cite{Subjakova:2020shi}. The detailed analysis of the curvature's effects on the phase diagram is ongoing. It will be presented elsewhere, while here we concentrate only on the aspects relevant to the position of the triple point. 

We wish to simultaneously inspect two finite limits of our matrix model, which zoom-in on different portions of the parameter space:
\begin{subequations}
\begin{eqnarray}
    \mathcal{S}(c_2,c_4,c_r) &=& \lim_{N\to\infty}\frac{\expval{S_N(c_2,c_4,c_r)}}{N^2}, 
   \\
   \widetilde{\mathcal{S}}(\widetilde{c}_2,\widetilde{c}_4,c_r) &=& \lim_{N\to\infty}\frac{\expval{S_N(\widetilde{c}_2,\widetilde{c}_4,c_r)}}{N^2}.
\end{eqnarray}
\end{subequations}
In a way, phase diagram of $\widetilde{\mathcal{S}}$ describes the structure of the infinity of the phase diagram of $\mathcal{S}$. $\mathcal{S}_0$ and $\widetilde{\mathcal{S}}_0$ will refer to $c_r=0$. We analyze $\mathcal{S}$ because it closely relates to $S_\text{GW}$ up to a light adjustment of coefficients (Appendix \ref{SNtoGW}). We also wish to complete the analysis of the  $\widetilde{\mathcal{S}}_0$ started in \cite{Prekrat:2020ptq}, which tells us about the scaling properties of the 3rd order $\updownarrow \to \uparrow\downarrow$ transition line. Notice that parameters' mass dimensions are:
\begin{equation}
    [c_2] = [c_4] = 2, 
    \qquad
    [c_r] = 0;
\end{equation}
therefore, both $c_2$ and $c_4$ in $\widetilde{\mathcal{S}}$ are chosen to scale the same way with the momentum cutoff $\Lambda \sim \sqrt{N}\mu $, whereas $c_r$ does not scale at all.

%%%%%%%%%%%%%%%%%%%%%%%%%%%%%%%%%%%%%%%%%%%%%%%%%%%%%%%%%%%%%%%%%%%%%%%%%%%%%%%%%%%%%
%%%%%%%%%%%%%%%%%%%%%%%%%%%%%%%%%%%%%%%%%%%%%%%%%%%%%%%%%%%%%%%%%%%%%%%%%%%%%%%%%%%%%
%%%%%%%%%%%%%%%%%%%%%%%%%%%%%%%%%%%%%%%%%%%%%%%%%%%%%%%%%%%%%%%%%%%%%%%%%%%%%%%%%%%%%

\section{$R$-off triple point}

In the spirit of Bayesian probability notation, we will write the coordinates $c_i$ of the triple point $T$ in the $R$-on and $R$-off case as $c_i(T|r)$ and $c_i(T|\slashed{r})$, respectively.

\begin{figure}[b]
\centering  
\includegraphics[scale=0.55]{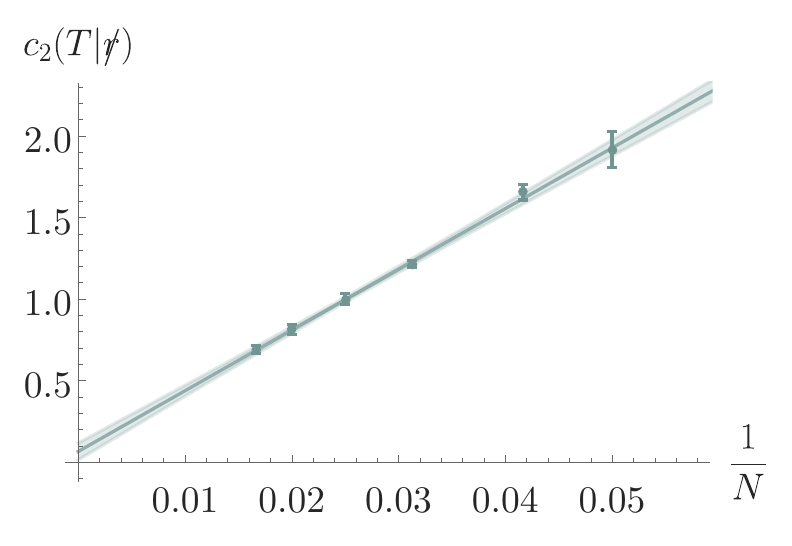}
\includegraphics[scale=0.55]{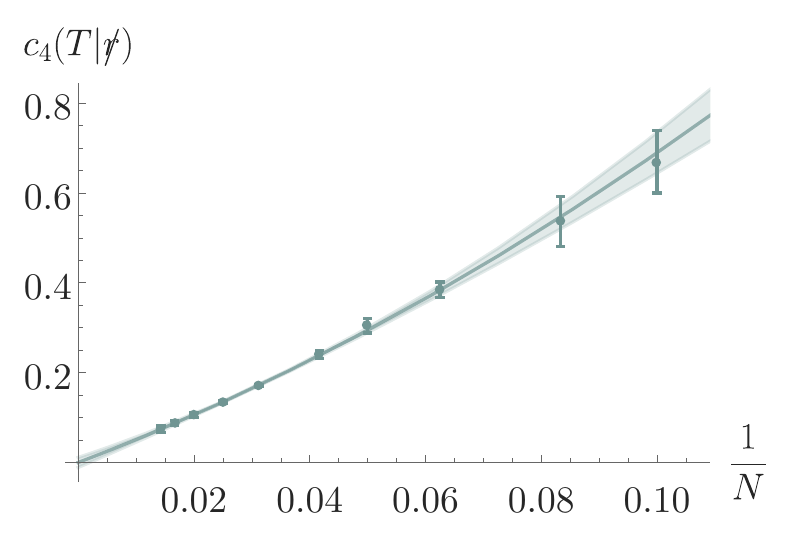}
\caption{Coordinates of the triple point in the $R$-off case as a function of the inverse matrix size.  (top)
Linear fit of the $c_2$-coordinate: $c_2(T|\slashed{r})=+0.07(5)+37(2)/N$. (bottom) Quadratic fit of the $c_4$-coordinate: $c_4(T|\slashed{r})=-0.000(12)+4.9(7)/N+20(10)/N^2$. Data gathered from susceptibility $\chi$ for $N\leq70$. The $\updownarrow \to \uparrow\uparrow$ transition line ending in $T$ shrinks with an increase in matrix size, and eventually disappears.}
\label{triple0}
\end{figure}

In \cite{Prekrat:2020ptq}, we found that the triple point of $\widetilde{\mathcal{S}}_0$ lies at $\widetilde{c}_4(T|\slashed{r})\lesssim 0.005$ (alternatively: $c_4(T|\slashed{r})\lesssim 0.14$ from $N=28$ data) and established the descending trend of $\widetilde{c}_4(T|\slashed{r})$ with increase in matrix size. In the meantime, we collected more data for matrix sizes up to $N=70$, allowing us to track the shrinking rate of the $\updownarrow \to \uparrow\uparrow$ transition line. Unexpectedly, this transition disappears entirely and the triple point collapses into the origin (FIG. \ref{triple0}).

Appendix \ref{proxies} provides details on locating the triple point from raw data and different attempted data fits (Table \ref{tab:fits}). We modeled small aberrations from the linear trend set by larger matrices by quadratic and power-law functions of $1/N$. All the estimates agree with triple point lying at the origin in the large $N$ limit, and the best one bounds its coordinates to 
\begin{equation}
    (c_2,c_4)_T\leq(0.16,0.018)
\end{equation} 
with the 95\% confidence level each, which is an order of magnitude improvement in precision.  

In addition, linear extrapolation of the slopes of transition lines for $N=24,32,40,50$ shows that they radiate from the triple point/origin as
\begin{subequations}
\begin{eqnarray}
    \updownarrow \to \uparrow\downarrow: \qquad c_2&=&7.1(8)c_4,
\\
    \uparrow\downarrow \to \uparrow\uparrow: \qquad c_2&=&17(1)c_4.
\end{eqnarray}
\end{subequations}
This is also how the phase diagram of $\widetilde{\mathcal{S}}_0$ looks like close to the origin, while away from it, its transition lines bend into $\sim\sqrt{\widetilde{c}_4}$.

It is important to notice that even if the triple point of $\mathcal{S}_0$ does not lie precisely at the origin, the triple point of $\widetilde{\mathcal{S}}_0$ will, due to $\widetilde{c}_i(T)=c_i(T)/N$. In fact, this holds for any alternate parameter rescaling $c_i/N^{\nu_i}$ by the positive power of the cutoff. This is in contrast with the $\phi^4$-model on the fuzzy sphere \cite{Kovacik:2018thy,Tekel:2017nzf}. The culprit could be in differing forms of the kinetic term, whose presence allows the $\updownarrow \to \uparrow\uparrow$ transition to develop in the first place.

%%%%%%%%%%%%%%%%%%%%%%%%%%%%%%%%%%%%%%%%%%%%%%%%%%%%%%%%%%%%%%%%%%%%%%%%%%%%%%%%%%%%%
%%%%%%%%%%%%%%%%%%%%%%%%%%%%%%%%%%%%%%%%%%%%%%%%%%%%%%%%%%%%%%%%%%%%%%%%%%%%%%%%%%%%%
%%%%%%%%%%%%%%%%%%%%%%%%%%%%%%%%%%%%%%%%%%%%%%%%%%%%%%%%%%%%%%%%%%%%%%%%%%%%%%%%%%%%%

\section{$R$-on shift and renormalization}

\begin{figure}
\centering  
\includegraphics[scale=0.55]{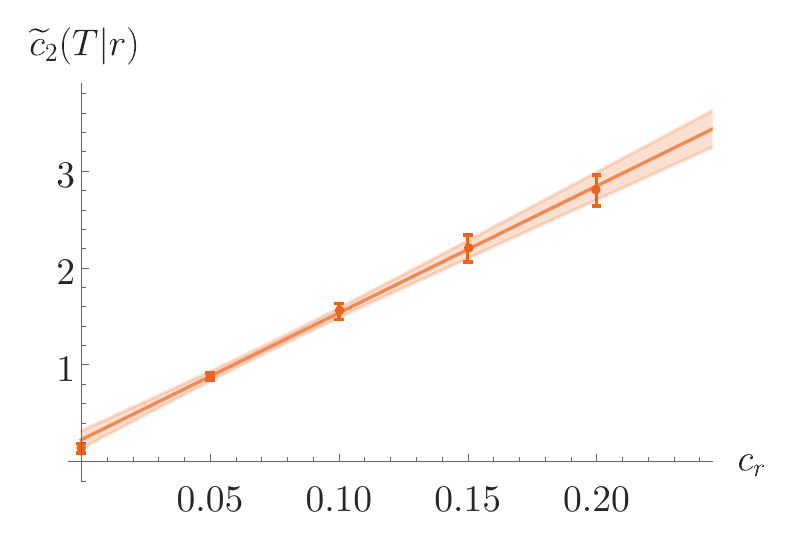}
\includegraphics[scale=0.55]{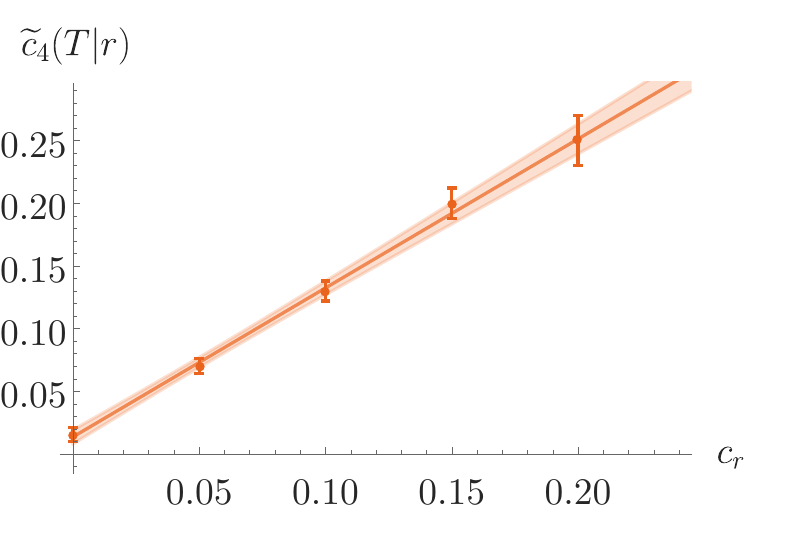}
\caption{Coordinates of the triple point in the $R$-on case as a function of the curvature coupling: $\widetilde{c}_2(T|r)=0.18(8)+15.5(7)c_r$, $\widetilde{c}_4(T|r)=0.014(6)+1.19(8)c_r$. Data gathered from specific heat $C$ for $N=24$.}
\label{tripleR}
\end{figure}

Coupling with curvature pushes the triple point---and with it the stripe phase---away from the origin proportionally to its strength (FIG. \ref{tripleR}). The simulated shift of the triple point in the $N=24$, $R$-on case 
\begin{equation}
    \widetilde{c}_2(T|r)=0.18(8)+15.5(7)c_r
\end{equation}
relative to the $R$-off value 
\begin{equation}
    \widetilde{c}_2(T|\slashed{r})=0.14(5),
\end{equation}
agrees well with the maximal prediction allowed by \eqref{shift}: 
\begin{equation}
    \max\delta\hspace{0.5pt}\widetilde{c}_2=\left(16-\frac{63}{2N}\right)c_r \approx 14.7c_r.
    \label{bound}
\end{equation}
The slight overshoot is discussed in Appendix \ref{proxies}. FIG. \ref{triple0} shows that the small intercept of $\widetilde{c}_2(T|r)$-line in FIG. \ref{tripleR} goes to zero with the increase in matrix size, so it is safe to assume that proportionality to $c_r$ becomes exact in the infinite size limit.

In the GW approach \cite{Grosse:2003nw}, renormalizability of the two-dimensional $\phi^4_\star$-model is assured by defining it as a $\Omega\to 0$ limit of the series of super-renormalizable models in which $\Omega$ itself does not renormalize and serves as a series label \cite{Wulkenhaar:habilitation2004}. It is chosen as
\begin{equation}
    \frac{1-\Omega^2}{1+\Omega^2}=
    \sqrt{1-\frac{1}{(1+\log(\Lambda/\Lambda_R))^2}},
\end{equation}
which for large cutoff $\Lambda \sim \sqrt{N}$ switches off as
\begin{equation}
    \Omega \sim \frac{1}{\log N}.
\end{equation}
Since Appendix \ref{SNtoGW} connects $\Omega$ and $c_r$ as
\begin{equation}
    c_r = \frac{\Omega^2/8}{1-\Omega^2/2}, 
\end{equation}
we consider the limit
where $c_r$ decreases as 
\begin{equation}
    c_r \sim \frac{1}{\log^2 N}.
\end{equation}
Combining this with $\delta\hspace{0.5pt}\widetilde{c}_2\propto c_r$ (that is: $\delta c_2\propto Nc_r$), would effectively swipe the stripe phase off to infinity as
\begin{equation}
    c_2(T) \sim \frac{N}{\log^2 N},
    \label{c2(T,N)}
\end{equation}
leaving the limiting model with a completely different phase diagram from the one obtained by simply setting $c_r=0$. 
 
Looking back at the equation of motion \eqref{EOM1} and its solutions \eqref{EOM2}, we see that curvature term prefers the trivial over striped vacuum. The action \eqref{S_N} also shows that the curvature itself compensates the attenuation of the coupling. Namely, for nearly ordered field configurations $\Phi^2\propto\1$, the curvature term dominates the potential by factor
\begin{equation}
    \frac{\Tr R\Phi^2}{\Tr\Phi^{2n}} \approx \frac{\Tr R}{\Tr\1} = -8N\left(1-\frac{31}{16N}\right) \sim N, 
\end{equation}
which multiplied by the coupling leads once more to ratio
\begin{equation}
   \frac{N}{\log^2 N}.
\end{equation}

It is instructive to also track the behaviour of the renormalized mass parameter as we turn off the curvature coupling. In \cite{Vinas:2014exa}, the divergent part of the mass renormalization in the $R$-on case is found to be:
\begin{equation}
    \delta m^2_\text{ren} = \frac{\lambda}{12\pi(1+\Omega^2)}\log\frac{\Lambda^2\theta}{\Omega},
\label{m2ren}
\end{equation}
which, adapted to our notation, gives the leading logarithmic mass divergence
\begin{equation}
    \delta c_2^\text{ren} \sim -\log N.
\end{equation}
Bare $c_2$ has to compensate this, increasing by $\abs{\delta c_2^\text{ren}}$. Since this grows slower than \eqref{c2(T,N)}, the bare $c_2$ required for the renormalization is located outside of the stripe phase---apparently, in the disordered phase. 

This differs from the $R$-off case with $T$ at the origin. Although we cannot directly set $\Omega=0$ in \eqref{m2ren}, we could try to turn off $\Omega$ much faster than in the GW approach. Taking, for example,
\begin{equation}
    \Omega \sim e^{-N},
\end{equation}
would give the leading divergence
\begin{equation}
    \delta m^2_\text{ren} \sim \lambda N, 
\end{equation}
and
\begin{eqnarray}
   \delta c_2(T) \sim Ne^{-2N} \to 0,
\end{eqnarray}
as expected in the $R$-off case. Thus, a suitably chosen infinitesimal $\lambda \sim 1/N$ would make the renormalization finite, leaving the bare mass inside the near-origin portion of the stripe phase for a range of physical mass choices in the perturbative regime.

%%%%%%%%%%%%%%%%%%%%%%%%%%%%%%%%%%%%%%%%%%%%%%%%%%%%%%%%%%%%%%%%%%%%%%%%%%%%%%%%%%%%%
%%%%%%%%%%%%%%%%%%%%%%%%%%%%%%%%%%%%%%%%%%%%%%%%%%%%%%%%%%%%%%%%%%%%%%%%%%%%%%%%%%%%%
%%%%%%%%%%%%%%%%%%%%%%%%%%%%%%%%%%%%%%%%%%%%%%%%%%%%%%%%%%%%%%%%%%%%%%%%%%%%%%%%%%%%%

\section{Concluding remarks}

This paper aimed to see if the renormalizable GW redefinition of the $\phi^4_\star$-model is reflected in its corresponding phase diagram and if it affects the extent of the stripe phase connected to the UV/IR mixing. With that in mind, we tracked the triple point position: this is where the stripe phase starts, spreading towards larger values of the mass and quartic parameters. We compared the matrix regularization of the $\phi^4_\star$-model with the disappearing curvature term to the one without one, since its inclusion is crucial for the renormalizability.  

We first refined the estimate \cite{Prekrat:2020ptq} of the triple point position when the curvature term is turned off and concluded that it collapses into the origin in the infinite matrix size limit, completely removing the $\updownarrow \to \uparrow\uparrow$ transition line.
As the curvature coupling turns on, it shifts the triple point towards larger values of the mass parameter, proportionally to its strength and the matrix size. GW coupling attenuation does not reverse this effect. Instead, the triple point moves to infinity, erasing the stripe phase.  

This leads to different phase diagrams for the renormalizable and non-renormalizable version of the $\phi^4_\star$-model. The former seems to be stripe phase-free, while in the latter, the stripe phase is tethered to the origin of the parameter space. This demonstrates that GW construction affects both renormalizability and phase structure.

We are currently completing the exploration of the $R$-on model, and we also plan to further inspect the $R$-off triple point for $N>70$ to make sure it lies at the origin. Although we observed the convincing curvature-mediated linear shift for $N=24$ in the $R$-on case, it would be prudent to confirm the effect for larger matrices as well. We would also like to simultaneously decrease the coupling and increase the matrix size since their combined effect was here indirectly deduced.

Additionally, we mean to revisit the GW-inspired gauge model \cite{Buric:2016lly} in the hope of uncovering the reverse effect: non-renormalizability due to retention of the stripe phase. After fixing the NC strength and scale, the model is left with only one adjustable parameter---the field strength coupling. Its additional stripe-like vacuum transforms into a trivial one for the infinite coupling, implying the phase diagram's possible structure: stripe phase for weak interaction and disordered phase for strong interaction. This agrees with the phase diagrams in FIG. \ref{diagrams}, where ordered phases lie at smaller quartic coupling compared to the disordered phase.

Another possible line of investigation could be the phase diagram of renormalizable spinor model on $\mathfrak{h}^\text{tr}$ \cite{Buric:2015vja} in context of fermionic matrix models \cite{Semenoff:1996vm,Arefeva:2019wjb}.

If this correspondence between renormalizability and phase structure proves to hold across models, it might be helpful when constructing new ones. For example, one could search numerically for early signatures of (non)renormalizability in their phase diagrams, assessing the new model's renormalizability potential, even before the involved and time-consuming analytical exploration.

%%%%%%%%%%%%%%%%%%%%%%%%%%%%%%%%%%%%%%%%%%%%%%%%%%%%%%%%%%%%%%%%%%%%%%%%%%%%%%%%%%%%%
%%%%%%%%%%%%%%%%%%%%%%%%%%%%%%%%%%%%%%%%%%%%%%%%%%%%%%%%%%%%%%%%%%%%%%%%%%%%%%%%%%%%%
%%%%%%%%%%%%%%%%%%%%%%%%%%%%%%%%%%%%%%%%%%%%%%%%%%%%%%%%%%%%%%%%%%%%%%%%%%%%%%%%%%%%%

%\vfill

\begin{acknowledgments}
This work was supported by the Serbian Ministry of Education, Science and Technological Development Grant ON171031 and by COST Action MP1405. The author would like to thank Maja Buri\'{c}, Denjoe O'Connor, Samuel Kov\'{a}\v{c}ik, and Sebasti\'an Franchino-Vi\~nas for valuable discussions and DIAS for hospitality and the financial support. 
\end{acknowledgments}

%%%%%%%%%%%%%%%%%%%%%%%%%%%%%%%%%%%%%%%%%%%%%%%%%%%%%%%%%%%%%%%%%%%%%%%%%%%%%%%%%%%%%
%%%%%%%%%%%%%%%%%%%%%%%%%%%%%%%%%%%%%%%%%%%%%%%%%%%%%%%%%%%%%%%%%%%%%%%%%%%%%%%%%%%%%
%%%%%%%%%%%%%%%%%%%%%%%%%%%%%%%%%%%%%%%%%%%%%%%%%%%%%%%%%%%%%%%%%%%%%%%%%%%%%%%%%%%%%

\appendix

\section{Model correspondence}
\label{SNtoGW}

According to \cite{Ydri:2016dmy}, mapping
\begin{equation}
    \phi \longleftrightarrow \Phi, 
    \qquad
    \int \longleftrightarrow \sqrt{\det2\pi\theta}\Tr
\end{equation}
connects field theory on Moyal space and matrix field theory with the same parameters.
Also, \cite{Buric:2009ss} provides a correspondence between $S_\text{GW}$ and $S_\mathfrak{h}$:
\begin{subequations}
\begin{eqnarray}
    S_\text{GW} &=& \left(1-\frac{\Omega^2}{2}\right) S_\mathfrak{h},
\\
    m^2 &=& \left(1-\frac{\Omega^2}{2}\right)\left(m_\mathfrak{h}^2-\frac{15}{2}\xi\mu^2\right),
\\
    \lambda &=&  \left(1-\frac{\Omega^2}{2}\right)\lambda_\mathfrak{h},
\\
    \Omega^2 &=&  8\epsilon^2\left(1-\frac{\Omega^2}{2}\right)\xi.
\end{eqnarray}
\end{subequations}
From these, by comparing \eqref{Sr} and \eqref{S_N}, it is easy to conclude that $S_\text{GW}$ and $S_N$ are connected by
\begin{subequations}
\begin{eqnarray}
    S_\text{GW} &=& \pi\left(1-\frac{\Omega^2}{2}\right) S_N,
\\
    m^2 &=& -\left(1-\frac{\Omega^2}{2}\right)\left(c_2+\frac{15}{2}c_r\right),
\\
    \lambda &=&  12\left(1-\frac{\Omega^2}{2}\right)c_4,
\\
    \Omega^2 &=&  8\left(1-\frac{\Omega^2}{2}\right)c_r,
\end{eqnarray}
\end{subequations}
in the large $N$ limit ($\theta^{12}=1/\mu^2$, units: $\mu=1$). 

Furthermore, action multiplier can be absorbed into the field during expectation value integration and will affect only $c_4$:
\begin{subequations}
\begin{eqnarray}
    \expval{\kappa S(c_2,c_4,c_r)}_{\kappa S} &=& \expval{S(c_2,c_4/\kappa,c_r)}_{S},\quad\;\;
\\
    \sqrt{\kappa}\expval{\Phi}_{\kappa S}(c_2,c_4,c_r) &=& \expval{\Phi}_{S}(c_2,c_4/\kappa,c_r),
\end{eqnarray}
\end{subequations}
yielding
\begin{subequations}
\begin{eqnarray}
    C_{\kappa S}(c_2,c_4,c_r) &=& C_S(c_2,c_4/\kappa,c_r),\;
\\
    \kappa\chi_{\kappa S}(c_2,c_4,c_r) &=& \chi_S(c_2,c_4/\kappa,c_r).
\end{eqnarray}
\end{subequations}
Since we are interested in the position of peaks of $C$ and $\chi$, this means that phase transition diagrams for $\kappa S$ and $S$ will be the same up to a reparametrization
\begin{equation}
    (c_2,c_4,c_r) \longleftrightarrow (c_2,c_4/\kappa,c_r).
\end{equation}
For phase diagrams of $S_\text{GW}$ and $\mathcal{S}$ in the $\Omega\to 0$ limit ($c_r\to 0$), this means:
\begin{equation}
    (m^2,\lambda) \longleftrightarrow (-c_2,12c_4/\pi).
\end{equation}

%%%%%%%%%%%%%%%%%%%%%%%%%%%%%%%%%%%%%%%%%%%%%%%%%%%%%%%%%%%%%%%%%%%%%%%%%%%%%%%%%%%%%
%%%%%%%%%%%%%%%%%%%%%%%%%%%%%%%%%%%%%%%%%%%%%%%%%%%%%%%%%%%%%%%%%%%%%%%%%%%%%%%%%%%%%
%%%%%%%%%%%%%%%%%%%%%%%%%%%%%%%%%%%%%%%%%%%%%%%%%%%%%%%%%%%%%%%%%%%%%%%%%%%%%%%%%%%%%

\section{Triple point proxies and fits}
\label{proxies}

\renewcommand{\arraystretch}{1.8}
\begin{table*}[!t]
\caption{\label{tab:fits} Different models of $R$-off triple point position fitting. All intercepts are consistent with the triple point located at the origin. A linear fit is performed for data subsets with higher $N$, where nonlinearities are imperceptible.}
\centering
\vspace{10pt}
\begin{tabular}{ccll}
    \hline\hline
    \;data\; &  \;\;model & $\qquad c_4(T|\slashed{r})$ fit & $c_2(T|\slashed{r})$ fit \\
    \hline 
    \multirow{3}{*}{$\chi$}  
    & \;\;linear  & 
    $\qquad c_4=-0.008(16)+5.7(7)/N$ &
    $c_2=+0.07(5)+37(2)/N$
    \\ \cline{2-4}
     & \;\;quadratic & 
    $\qquad c_4=-0.000(11)+4.9(7)/N+20(10)/N^2\qquad$ &
    $c_2=+0.04(8)+39(5)/N-42(60)/N^2$\;
    \\ \cline{2-4}
    & \;\;power law  & 
    $\qquad c_4=+0.015(15)+12(4)/N^{1.24(11)}$ &
    $c_2=-0.0(2)+31(10)/N^{0.93(14)}$
    \\
    \hline
    $C$ 
    & \;\;linear  & 
    $\qquad c_4=-0.12(10)+16(3)/N$ &
    $c_2=+0.9(9)+93(21)/N$
    \\
    \hline\hline
\end{tabular}
\end{table*}

We here discuss triple point proxies used for FIG. \ref{triple0} and FIG. \ref{tripleR}. 

\begin{figure}[!b]
\centering  
\includegraphics[scale=0.53]{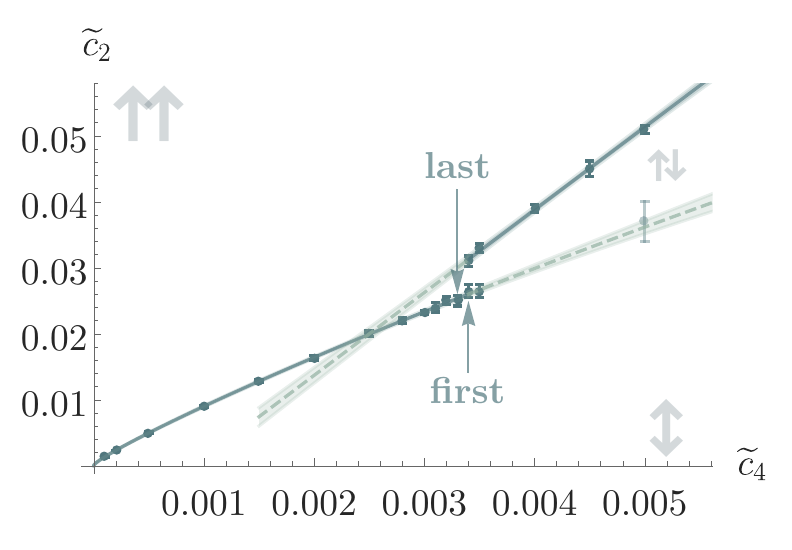}
\includegraphics[scale=0.51]{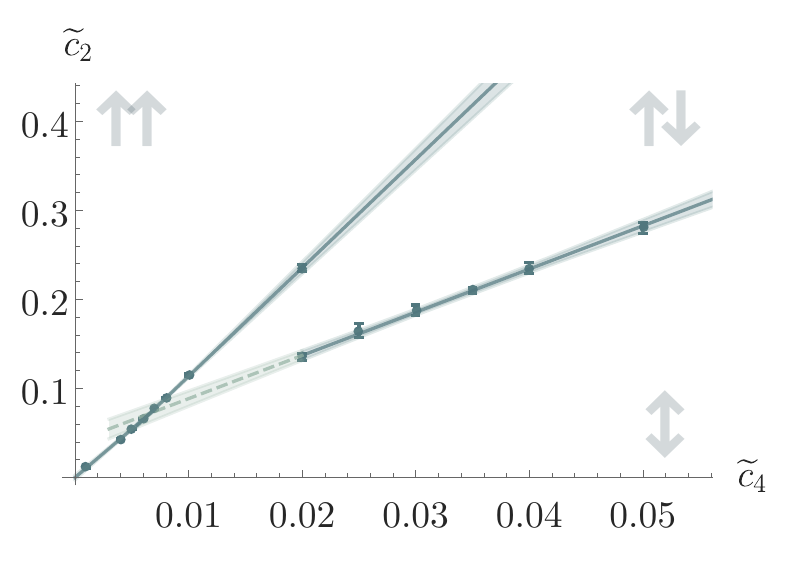}
\caption{Phase diagrams for $N=40$ in the vicinity of the triple point. (top) Change in slope of the $\uparrow\uparrow$-phase border indicates a triple point. Arrows mark the last point on $\updownarrow \to \uparrow\uparrow$ and the first point on $\updownarrow \to \uparrow\downarrow$ lines.  Constructed from $\chi$-data. (bottom) Extrapolated transition lines with 83\% confidence intervals. Constructed from $C$-data.}
\label{N40}
%\vspace{-30pt}
\end{figure}

The split of the profile of $\chi$ into two separate peaks as the $\updownarrow \to \uparrow\uparrow$ line bifurcates into $\updownarrow \to \uparrow\downarrow$ and $\uparrow\downarrow \to \uparrow\uparrow$, changes the slope of the $\uparrow\uparrow$-phase border $\partial_{\,\uparrow\uparrow}$ in FIG. \ref{N40}. Therefore, the midpoint between the last point in the one-peak and the first point in the two-peak regime served as the triple point proxy for FIG. \ref{triple0}. We used the standard deviation of the triangular distribution ending at these two points as the triple point position uncertainty. For $N\leq24$, the two peaks are not completely separated in the triple point region, so we instead took the intersection of extrapolated transition lines.

For consistency, we also checked the $C$-data, which has less predictive power due to larger uncertainties and distance from the triple point region. We there extrapolated $\updownarrow \to \uparrow\downarrow$ transition line to its intersection with $\partial_{\,\uparrow\uparrow}$. To get 68\% confidence intervals of the intersection point coordinates, we used 83\% confidence intervals of transition line fits, since the probability of triple point belonging to their intersection is given by
\begin{equation}
    P(\updownarrow \to \uparrow\downarrow \cap \;\partial_{\,\uparrow\uparrow})=P(\updownarrow \to \uparrow\downarrow)P(\partial_{\,\uparrow\uparrow})
\end{equation}
and $0.68\approx0.83^2$.

\begin{figure}[!t] 
\centering  
\includegraphics[scale=0.51]{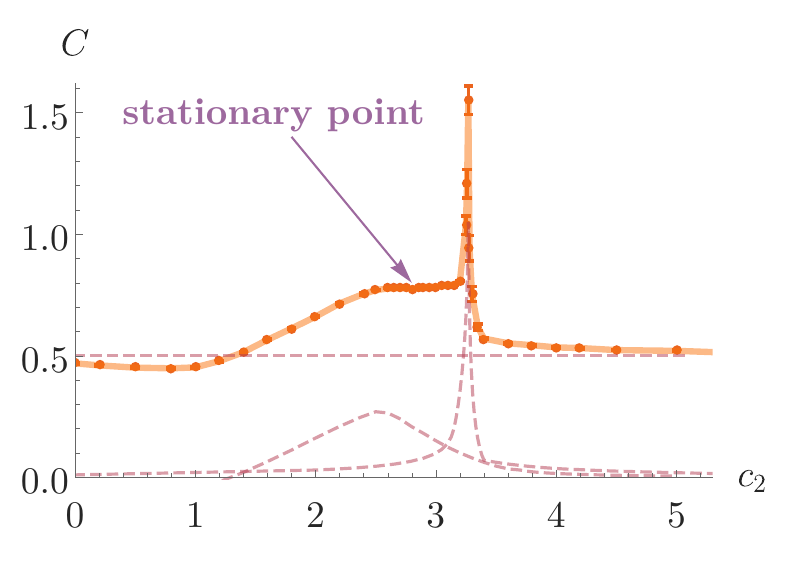}
\includegraphics[scale=0.51]{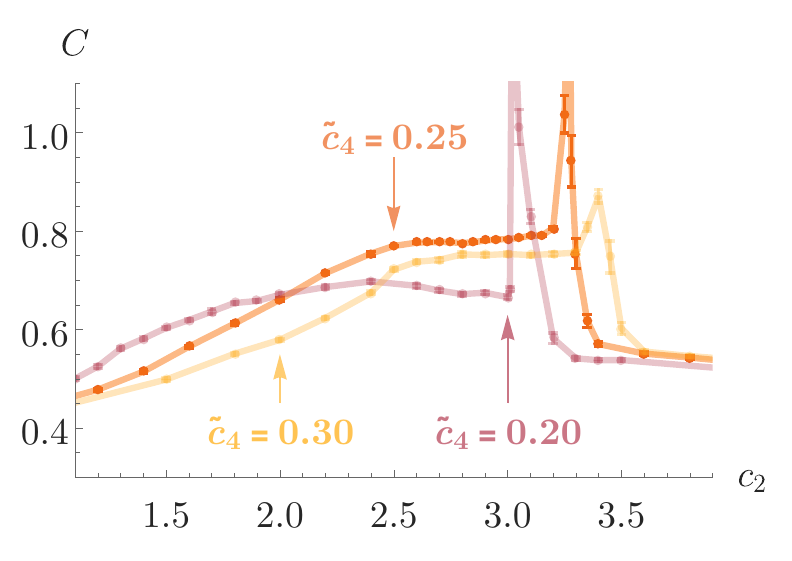}
\caption{(top) Triple point region for $N=24$, $c_r=0.2$ at $\widetilde{c}_4=0.25$ resolved into two peaks. The stationary point at the plateau is chosen as the triple point proxy. (bottom) Plateau at $\widetilde{c}_4=0.25$ mounts above plateaus at $\widetilde{c}_4=0.20$ and $\widetilde{c}_4=0.30$, building a wall between phases.}
\label{Ctriple}
\end{figure}

In the $R$-on case, we used contour diagrams (e.g. FIG. \ref{diagrams}) to detect the beginning of the $\uparrow\downarrow$-phase from $C$-data. We looked at the bright triple point-peak position and then checked the neighboring raw data to pinpoint its exact location. As it turns out, the peak resolves into two very closely spaced convoluted peaks---which presumably coincide when matrix size increases---joined by a wall that separates phases (bottom of FIG. \ref{Ctriple}). For FIG. \ref{tripleR}, we used the position of the protruding peak shown in FIG. \ref{Ctriple}, which gave a slightly higher estimate for the slope of the $c_2(T|r)=f(c_r)$ line than \eqref{bound}. The stationary point on this wall seems a more realistic estimator of the triple point position, but it is also more difficult to measure. A rough estimate using the stationary point
\begin{equation}
    \widetilde{c}_2(T|r)=13.2(11)c_r+0.24(9),
\end{equation}
fits within the interval \eqref{shift} and is close to its upper bound.
For comparison, the slope calculated from the smaller peaks is around 12.  

Looking at FIG. \ref{diagrams}, we see a small oval local minimum region with a bright triple point peak at its lower left edge. Eigenvalue distribution there has the characteristics of the $\uparrow\downarrow$-phase. We do not believe this to constitute a separate phase but a finite-size effect that collapses into a triple point as the matrix size increases. This should be, of course, checked at larger $N$. In addition, between the triple point and origin in the $R$-on case, there is a transitional region where curvature eigenvalues in \eqref{phiR} slowly activate as we go from $\updownarrow$ to $\uparrow\uparrow$-phase. This might constitute a separate partially-ordered phase, but requires more data and further analysis. 

Different extrapolations of the triple point position as a function of the inverse matrix size are collected in the TABLE \ref{tab:fits}.

\vfill

%%%%%%%%%%%%%%%%%%%%%%%%%%%%%%%%%%%%%%%%%%%%%%%%%%%%%%%%%%%%%%%%%%%%%%%%%%%%%%%%%%%%%
%%%%%%%%%%%%%%%%%%%%%%%%%%%%%%%%%%%%%%%%%%%%%%%%%%%%%%%%%%%%%%%%%%%%%%%%%%%%%%%%%%%%%
%%%%%%%%%%%%%%%%%%%%%%%%%%%%%%%%%%%%%%%%%%%%%%%%%%%%%%%%%%%%%%%%%%%%%%%%%%%%%%%%%%%%%

%\bibliographystyle{h-physrev}
\bibliography{paper.bib}

\end{document}